\documentclass[%
 reprint,
superscriptaddress,
 amsmath,amssymb,
 aps,
floatfix,
]{revtex4-1}

\usepackage{color}

\usepackage{graphicx}
\usepackage{dcolumn}
\usepackage{bm}
\usepackage{mhchem}
\usepackage{nicefrac}

\newcommand{\rbna}{\ce{Rb2NaTi3F12}}
\newcommand{\csna}{Cs$_2$NaTi$_3$F$_{12}$}
\newcommand{\csk}{Cs$_2$KTi$_3$F$_{12}$}

\begin{document}

\title{From kagome strip to kagome lattice: Realizations of frustrated $S=\nicefrac{1}{2}$ antiferromagnets in Ti(III) fluorides}

\author{Harald O. Jeschke}
\affiliation{Research Institute for Interdisciplinary Science, Okayama University, Okayama 700-8530, Japan}

\author{Hiroki Nakano}
\affiliation{Graduate School of Material Science, University of Hyogo, Kamigori, Hyogo 678-1297, Japan}

\author{T\^oru Sakai}
 \affiliation{Graduate School of Material Science, University of Hyogo, Kamigori, Hyogo 678-1297, Japan}
\affiliation{National Institutes for Quantum and Radiological Science and Technology (QST), SPring-8, Sayo, Hyogo 679-5148, Japan}

\date{\today}

\begin{abstract}
We investigate the connection between highly frustrated kagome based Hamiltonians and a recently synthesized family of materials containing Ti$^{3+}$ $S=\nicefrac{1}{2}$ ions. Employing a combination of all electron density functional theory and numerical diagonalization techniques, we establish the Heisenberg Hamiltonians for the distorted kagome antiferromagnets {\rbna}, {\csna} and {\csk}. We determine magnetization curves in excellent agreement with experimental observations. Our calculations successfully clarify the relationship between the experimental observations and the magnetization-plateau behavior at $\nicefrac{1}{3}$ height of the saturation and predict characteristic behaviors under fields that are higher than the experimentally measured region. We demonstrate that the studied Ti(III) family of materials interpolates between kagome strip and kagome lattice.
\end{abstract}


\maketitle

{\bf Introduction.-} Quantum antiferromagnets on the kagome lattice have fascinated experimental and theoretical physicists for a long time; in particular since the synthesis of high quality samples of herbertsmithite~\cite{Shores2005}, the search for quantum spin liquid candidates in highly frustrated kagome lattice materials has intensified~\cite{Lee2008,Norman2016}. Cu$^{2+}$ based materials like \ce{ZnCu3(OH)6Cl2} (herbertsmithite) or \ce{BaCu3V2O8(OH)2} (vesignieite)~\cite{Okamoto2009} or the vanadium oxyfluoride \ce{[NH4]2[C7H14N][V7O6F18]}~\cite{Aidoudi2011} have been at the forefront of the discussion. There are also effects of strong magnetic frustration and unconventional behavior in imperfect kagome lattices like \ce{ZnCu3(OH)6SO4}\cite{Li2014}  or breathing kagome lattices.
Recently, an interesting new family of $S=\nicefrac{1}{2}$ kagome materials has been realized based on magnetic Ti$^{3+}$ ions; three compounds {\rbna}, {\csna} and {\csk} have been reported\cite{Goto2016}. They are new members of a large class of materials; besides several members involving Cu$^{2+}$ like the possible valence bond solid \ce{Rb2Cu3SnF12}~\cite{Matan2010} there are $S=2$ Mn$^{3+}$ based members like \ce{Cs2LiMn3F12}~\cite{Englich1997} and the more recent $S=1$ V$^{3+}$ based \ce{Cs2KV3F12}~\cite{Goto2017} or $S=\nicefrac{3}{2}$ Cr$^{3+}$ containing \ce{Cs2KCr3F12}~\cite{Goto2018}. The Ti$^{3+}$ based materials are particularly attractive because spin orbit coupling is expected to be much smaller than in Cu$^{2+}$; thus, in the new $S=\nicefrac{1}{2}$ materials, the Dzyaloshinsky-Moriya interaction which complicates the discussion of many Cu$^{2+}$ based frustrated magnets~\cite{Mendels2010} promises to be much less important.

Theoretically, the $S=\nicefrac{1}{2}$ kagome lattice antiferromagnet has been studied intensively, using density-matrix renormalization group (DMRG)~\cite{Depenbrock2012}, numerical diagonalization~\cite{Zeng1990,HN_TSakai_kgm_1_3_JPSJ2014}, series expansion methods~\cite{Singh1992}, bosonization~\cite{Schnyder2008} and many other techniques. One possible strong distortion of the kagome lattice leads to kagome strips~\cite{Shimokawa2013} which have been discussed as $\Delta$ chains~\cite{Nakamura1996} or sawtooth lattice~\cite{Blundell2003} for many years. Recently, kagome strips were studied using the DMRG technique~\cite{Morita2018}. It is interesting to look for approximate realizations of kagome strips in real materials. Crystal structures in which the symmetry of the kagome lattice is broken by an orthorhombic distortion are possible candidates for the realization of kagome strips; however, which model is actually realized needs to be discussed at the level of electronic structure rather than crystal structure only.

\begin{figure}[tb]
\includegraphics[width=0.85\linewidth]{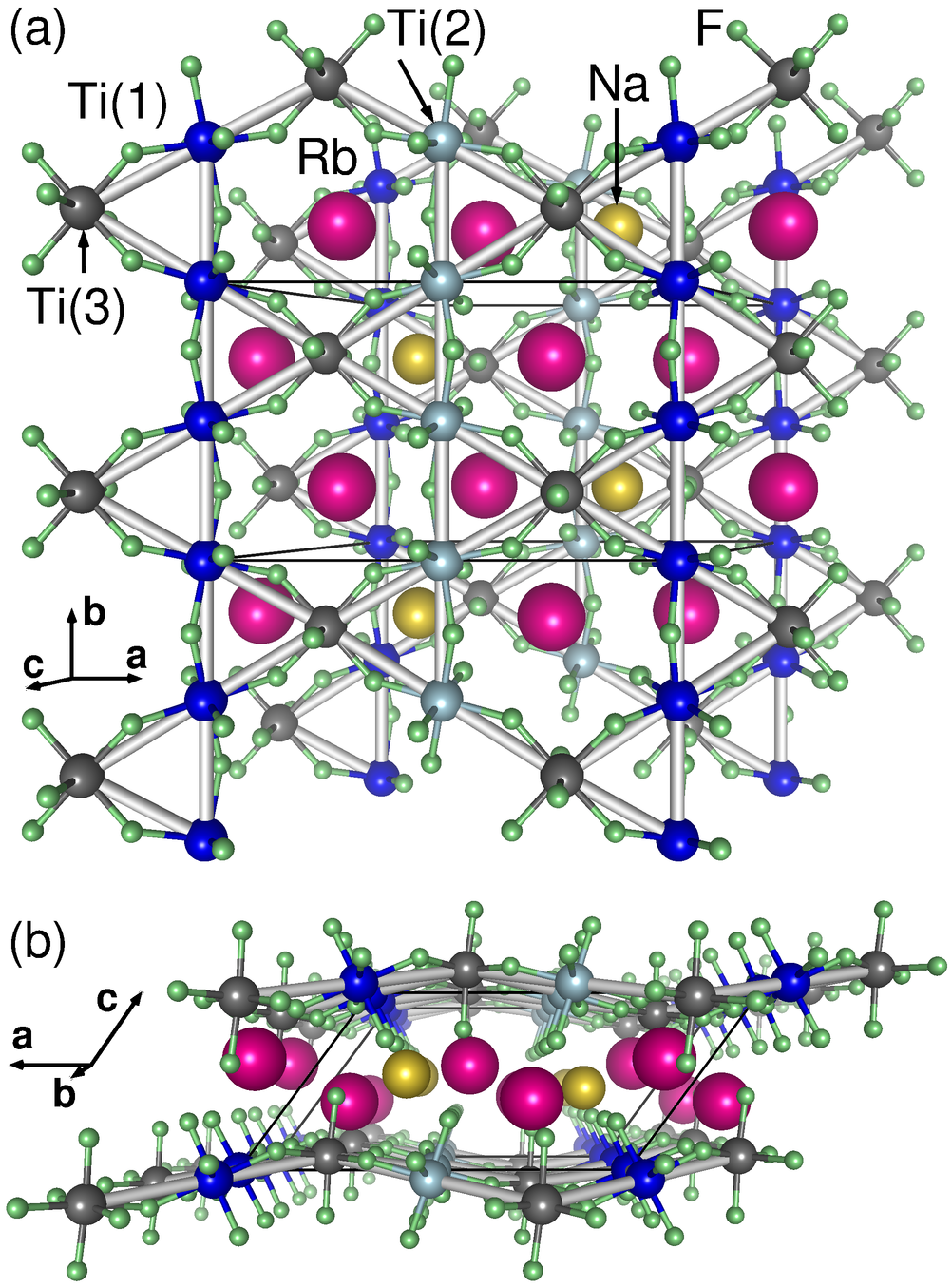}
\caption{(Color online) Crystal structure of {\rbna} with $P\,2_1/m$ space group~\protect\cite{Goto2016}. The kagome lattice formed by the three symmetry inequivalent Ti$^{3+}$ ions Ti(1), Ti(2) and Ti(3) is distorted and buckled.  }
\label{fig:structure}
\end{figure}

In this work, we address the problem that while the new $S=\nicefrac{1}{2}$ materials {\rbna}, {\csna} and {\csk} have been characterized structurally and magnetically, their Hamiltonian is essentially unknown. We will apply energy mapping techniques to evaluate the Heisenberg Hamiltonian up to third nearest neighbours in the distorted kagome lattice. We find that the three materials interpolate between a nearly pure $\Delta$ chain behaviour and an only slightly distorted kagome lattice behaviour. Using numerical diagonalization of clusters with up to 36 sites, we obtain excellent agreement with the measured magnetization curves.

\begin{figure*}[ht]
\includegraphics[width=0.98\textwidth]{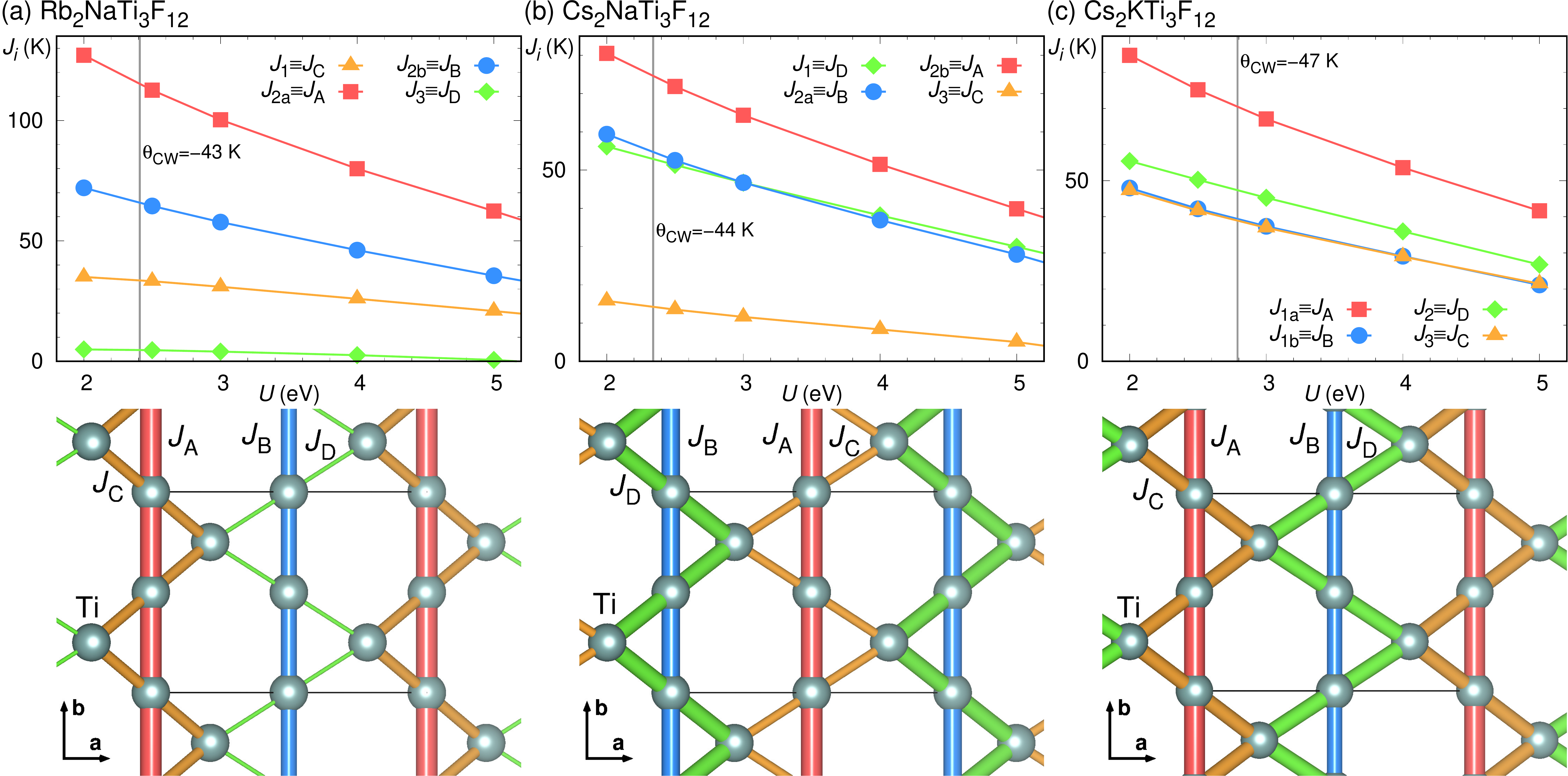}
\caption{Heisenberg Hamiltonian parameters for (a) {\rbna}, (b) {\csna} and (c) {\csk} obtained from fits to DFT total energies using a GGA+U functional for Ti $3d$ at  $J_{\rm H}=0.64$~eV as function of $U$. Vertical lines indicates the $U$ value at which the couplings reproduce
  the experimental Curie-Weiss temperatures $\Theta_{\rm CW}$. $J_1$ to $J_3$ label exchange paths geometrically according to increasing Ti-Ti distance $d_{\rm Ti-Ti}$ . $J_{2a}$/$J_{2b}$ distinguish paths with identical $d_{\rm Ti-Ti}$ but symmetry inequivalent Ti environment. $J_{\rm A}$/$J_{\rm C}$ labels the kagome strip belonging to the largest exchange coupling $J_{\rm A}$, $J_{\rm B}$/$J_{\rm D}$ labels the other, symmetry inequivalent kagome strip in the unit cell. The structures in the lower panels represent the geometry and the topology; bond cross-sectional areas are proportional to the size of the couplings. A clear evolution from kagome strip plus 1D chain in {\rbna} to slightly distorted 2D kagome model in {\csk} is observed. } \label{fig:couplings}
\end{figure*}

\begin{table}[hbt]
\caption{(Color online) Exchange couplings for {\rbna}, {\csna} and {\csk} determined by energy mapping, using the exchange correlation function GGA+U with $U$ values that reproduce the experimental Curie-Weiss temperatures $\Theta_{\rm CW}=-43$~K,  $\Theta_{\rm CW}=-44$~K  and $\Theta_{\rm CW}=-47$~K, respectively. See Figure~\ref{fig:couplings} for the assignment of $J_{\rm A}$, $J_{\rm B}$, $J_{\rm C}$ and  $J_{\rm D}$.}\label{tab:couplings}
\begin{ruledtabular}
\begin{tabular}{cllll}
 material &               $J_{\rm A}$~(K)&  $J_{\rm B}/J_{\rm A}$&  $J_{\rm C}/J_{\rm A}$&  $J_{\rm D}/J_{\rm A}$\\\hline
 {\rbna}      &115.0 & 0.572&  0.292&  0.040\\
  {\csna}       &74.6  & 0.733 & 0.190 & 0.709\\
 {\csk}       & 70.4  & 0.558&   0.552  & 0.673\\
\end{tabular}
\end{ruledtabular}
\end{table}

{\bf Methods.-} We calculate electronic structure and total energies with the full potential local orbital (FPLO) basis set~\cite{Koepernik1999} and the GGA functional~\cite{Perdew1996}.  Strong electronic correlations in Ti $3d$ orbitals are treated with the GGA+U~\cite{Liechtenstein1995} exchange and correlation functional. Hunds rule coupling for Ti $3d$ was fixed at $J_H=0.64$~eV~\cite{Mizokawa1996}. Heisenberg Hamiltonian parameters are extracted by mapping total energies for 15 different spin configurations to six exchange interactions~\cite{Jeschke2011,Iqbal2015,Guterding2016}.

To obtain magnetization processes, we also carry out numerical diagonalizations of finite-size clusters with Heisenberg interactions determined from the density functional theory calculations. Our numerical diagonalizations are performed based on the Lanczos and/or Householder algorithms in the subspace belonging to $\sum _j S_j^z=M_{z}$, where  the $z$-axis is taken as the quantized axis of each $S=\nicefrac{1}{2}$ spin operator ${\bf S}_j$ at site $j$. Our numerical diagonalizations give the lowest energy of the Heisenberg Hamiltonian in the subspace characterized by $M_{z}$, which leads to the magnetization curve. (See details in the Appendix~\ref{app:B}.) Some of the Lanczos diagonalizations were carried out using MPI-parallelized code that was originally developed in the study of Haldane gaps~\cite{HNakano_HaldaneGap_JPSJ2009}. The usefulness of our program was confirmed in large-scale parallelized calculations~\cite{HNakano_kgm_gap_JPSJ2011,HNakano_s1tri_LRO_JPSJ2013,HN_TSakai_kgm_1_3_JPSJ2014,HN_TSakai_kgm_S_JPSJ2015,HN_YHasegawa_TSakai_dist_shuriken_JPSJ2015,HN_TSakai_dist_tri_JPSJ2017,HN_TSakai_tri_NN_JPSJ2017,HN_TSakai_kgm45_JPSJ2018,YHasegawa_HN_TSakai_dist_shuriken_PRB2018,TSakai_HN_ICM018,HN_TSakai_S2HaldaneGap_JPSJ2018,HN_TSakai_OrthDimer_JPSJ2018}.

{\bf Results.-} Our calculations are based on the structures of isostructural {\rbna} (shown in Figure~\ref{fig:structure}), {\csna} and {\csk} as determined by Goto {\it et al.}~\cite{Goto2016}. Two Ti$^{3+}$ ion chains running along the crystallographic $b$ axis are different by symmetry. Geometrically, the kagome lattices formed by Ti(1), Ti(2) and Ti(3) sites are only about 1{\%} distorted. However, the decisive factor for the magnetic properties is the electronic distortion, revealed by the result of the energy mapping, Figure~\ref{fig:couplings}. The top of the figure shows the exchange couplings calculated by fitting total energies obtained with the GGA+U exchange correlation functional as function of the onsite correlation strength $U$. We are fitting to the Heisenberg Hamiltonian in the form 
\begin{equation}
  {\cal H}=\sum_{i<j} J_{ij} {\bf S}_i\cdot {\bf S}_j\,.
\label{eq:H}
\end{equation} 
Total moments are exact multiples of 1~$\mu_{\rm B}$. The energies of the fifteen considered spin configurations can be fitted extremely well (see also Appendix~\ref{app:A}), and statistical errors bars are smaller than the size of the symbols. Grey vertical lines indicate the interpolated $U$ value at which the set of couplings matches the experimental Curie-Weiss temperature as determined in Ref.~\onlinecite{Goto2016}. These $U$ values are in the range 2.3~eV to 2.8~eV which is reasonable for Ti. These sets of exchange couplings are listed in Table~\ref{tab:couplings}. The lower part of Figure~\ref{fig:couplings} illustrates the obtained Hamiltonians by representing the relative strength of the couplings as cross-sectional area of the Ti-Ti bonds. The first major result is the observation that the three considered materials realize three quite different Hamiltonians even though they are very similar structurally. The Hamiltonian of {\rbna} is dominated by an anisotropic kagome strip formed by Ti(1) ($J_{\rm A}$) and Ti(3) ($J_{\rm C} \approx 0.3 J_{\rm A}$) and a one-dimensional chain of Ti(2) ($J_{\rm B}$). In {\csna}, we have an almost isotropic kagome strip formed by Ti(2) and Ti(3) ions ($J_{\rm B} \approx J_{\rm D}$) and, nearly decoupled from that, a one-dimensional chain formed by the Ti(1) ions ($J_{\rm B}$). Finally, while in {\csk} the one-dimensional coupling of Ti(1) ions ($J_{\rm A}$) is a bit larger than the other three in-plane couplings, all other couplings in the kagome plane ($J_{\rm B}\approx J_{\rm C} \approx J_{\rm D}$) are also substantial, making {\csk} a frustrated two-dimensional antiferromagnet. We also determined two selected interlayer couplings corresponding to fifth and tenth nearest neighbor Ti-Ti distances. They are so small that they can be neglected in the further discussion.

\begin{figure}[tb]
\includegraphics[width=\linewidth]{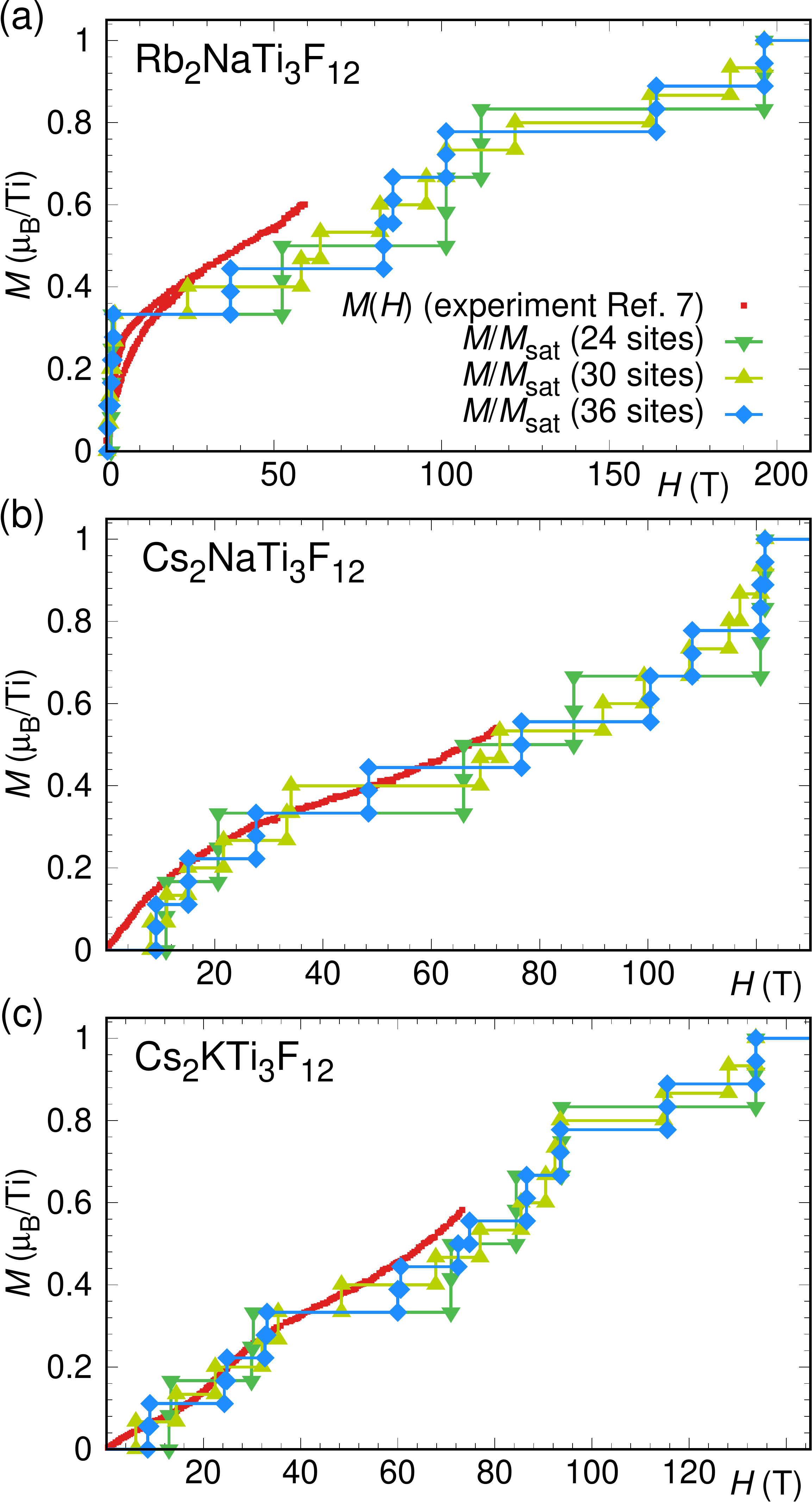}
\caption{(Color online) Comparison between experimental and calculated magnetization curves for all three compounds. The experimental data are from Ref.~\cite{Goto2016}.}
\label{fig:magstep}
\end{figure}

We now connect the Hamiltonians to the magnetic measurements by carrying out numerical diagonalization calculations for $N=24$, $N=30$ and $N=36$ site clusters.
We show $M(H)$ curves for {\rbna}, {\csna} and {\csk} in Figures~\ref{fig:magstep}~(a)-(c). The theoretical magnetic field $h$ given in units of $J_{\rm A}$ is converted to $H$ in Tesla using $H=h J_{\rm A}/2/0.6717$~T/K. The factor $1/2$ is introduced because the DFT determination of the $J_i$ in the Hamiltonian \eqref{eq:H} is done without double counting of bonds. The experimental data points in Figure~\ref{fig:magstep} are taken from Ref.~\onlinecite{Goto2016}.
Theoretical curves successfully capture characteristics of the experimental observations in spite of the limitation that our theoretical results have finite resolution due to the size of the computationally accessible clusters. It is, in particular, clarified that our results at and around the $\nicefrac{1}{3}$ height of the saturation can explain the behavior of the measured $M(H)$ curve for each material, as discussed in detail in the following. 

{\bf Discussion.-} A marked feature of the reported experimental result for \ce{Rb2NaTi3F12} shown in Fig.\ref{fig:magstep}~(a) is that there appears a considerably large gradient in the $M(H)$ curve from $M=0$ to approximately $M\sim 0.2$; above $M\sim 0.3$, on the other hand, the $M(H)$ curve shows relatively smaller gradients. Our numerical-diagonalization results reproduce well a very steep gradient below $M=\nicefrac{1}{3}$ and a contrastively small gradient around $M\sim 0.4$ above $M=\nicefrac{1}{3}$. Our results suggest that the saturation field is $H\sim 200$~T, which is much larger than the experimentally measured region. Note also that our results predict that a plateau-like behavior appears at $M\sim 0.8$ from approximately 100~T to a field near the suggested saturation. 

In Fig.~\ref{fig:magstep}~(b) for \ce{Cs2NaTi3F12}, the experimental data shows that the gradient at $M=0$ is larger than gradients around $M\sim 0.4$ but smaller than the steep gradient of \ce{Rb2NaTi3F12} at $M=0$. Our numerical-diagonalization results, especially around $M=\nicefrac{1}{3}$, capture well these behaviors. Our results suggest that the saturation field is $H\sim 120$~T. Although the height at $M=\nicefrac{2}{3}$ seems like a plateau for $N=24$, this behavior is considered to be a finite-size phenomenon because the widths for $N=30$ and 36 becomes much smaller than that for $N=24$. 

Ref.~\onlinecite{Goto2016} reported that {\csk} shows a concave behavior in the region approximately from 10~T to 20~T in its $M(H)$ curve as well as a peak at 23~T in its $dM/dH$. The peak leads to an abrupt increase of magnetization.
$dM/dH$ for {\csk} in Ref.~\onlinecite{Goto2016} reveals 
an inflection point approximately at 30~T. 
Since an inflection point in $dM/dH$ corresponds 
to an edge of a specific plateau at a nonzero temperature, 
the inflection point at 30~T can be considered to 
correspond to the edge of the $M=\nicefrac{1}{3}$ plateau 
on the low-field side. 
Our theoretical results for the edge are observed around 30-36~T;  
thus, the experimental and theoretical results agree well with each other.
Our theoretical result for $N=36$ shows a plateau-like behavior with a relatively wider region at $M=\nicefrac{1}{9}$ with a rapid increase of $M$ at the higher-field edge. Our theoretical results are very similar to the experimental observation although it is unclear at present whether or not the behavior of $M=\nicefrac{1}{9}$ corresponds to a spin-gapped behavior. Our theoretical results also suggest that the saturation field is $H\sim 130$~T and that there appears a plateau-like behavior at $M\sim 0.8$ from approximately 90~T to a field near the suggested saturation. 
Recently, Ref.~\onlinecite{Shirakami_JPSmeeting2018autumn} reported that 
the measurements of another material Cs$_2$LiTi$_3$F$_{12}$, 
which is a member of the same family of distorted kagome systems, 
show a behavior similar to that of {\csk}. 
In the magnetization curve of Cs$_2$LiTi$_3$F$_{12}$, 
a plateau-like behavior with height lower than $M=1/3$ 
accompanied by the $M=1/3$ plateau is observed~\cite{Yoshimuraprivate}. 
Investigation of the magnetization curve of the new systems with the methods of the present work will be an interesting future study, strengthening the Ti(III) fluorides as a platform for diverse distorted $S=1/2$ kagome systems.

{\bf Conclusions.-} We study three kagome-strip materials, \ce{Cs2KV3F12}, \ce{Rb2NaTi3F12}, and \ce{Cs2NaTi3F12} by density functional theory and numerical-diagonalization calculations.
The $S=\nicefrac{1}{2}$ Hamiltonians revealed by density functional theory based energy mapping indicate that the changes in the alkali metal spacers tune the materials between one- and two-dimensional magnetic behavior. We can show that \ce{Rb2NaTi3F12} approximately realizes an anisotropic $\Delta$ chain, and \ce{Cs2NaTi3F12} is close to an isotropic $\Delta$ chain. Thus, we have demonstrated that these two materials are very close to realizing important one-dimensional quantum spin systems. 
Our theoretical results concerning the magnetization curves agree well with experimental results reported in Ref.~\onlinecite{Goto2016}. 
Since the present systems reveal a quasi-one-dimensionality 
in the two-dimensional lattices, the DMRG technique, 
which was successfully used in Refs.~\onlinecite{Shimokawa2013} 
and \onlinecite{Morita2018} on related problems, 
could provide us with valuable information on the systems. 
By predicting the behaviour at higher magnetic fields, we hope to inspire further experimental studies.

\acknowledgments
We wish to thank Prof. Kazuyoshi Yoshimura for fruitful discussions. 
This work was partly supported by JSPS KAKENHI Grant Numbers 16K05418, 16K05419, 16H01080 (JPhysics), and 18H04330 (JPhysics). Nonhybrid thread-parallel calculations in numerical diagonalizations were based on TITPACK version 2 coded by H. Nishimori. In this research, we used the computational resources of Fujitsu PRIMERGY CX600M1/CX1640M1(Oakforest-PACS) provided by Joint Center for Advanced High Performance Computing through the HPCI System Research project (Project ID: hp180053). Some of the computations were performed using the facilities of the Department of Simulation Science, National Institute for Fusion Science; Institute for Solid State Physics, The University of Tokyo; and Supercomputing Division, Information Technology Center, The University of Tokyo. 

\appendix 

\section{Energy mapping}\label{app:A}

\begin{figure}[tb]
  \includegraphics[width=\linewidth]{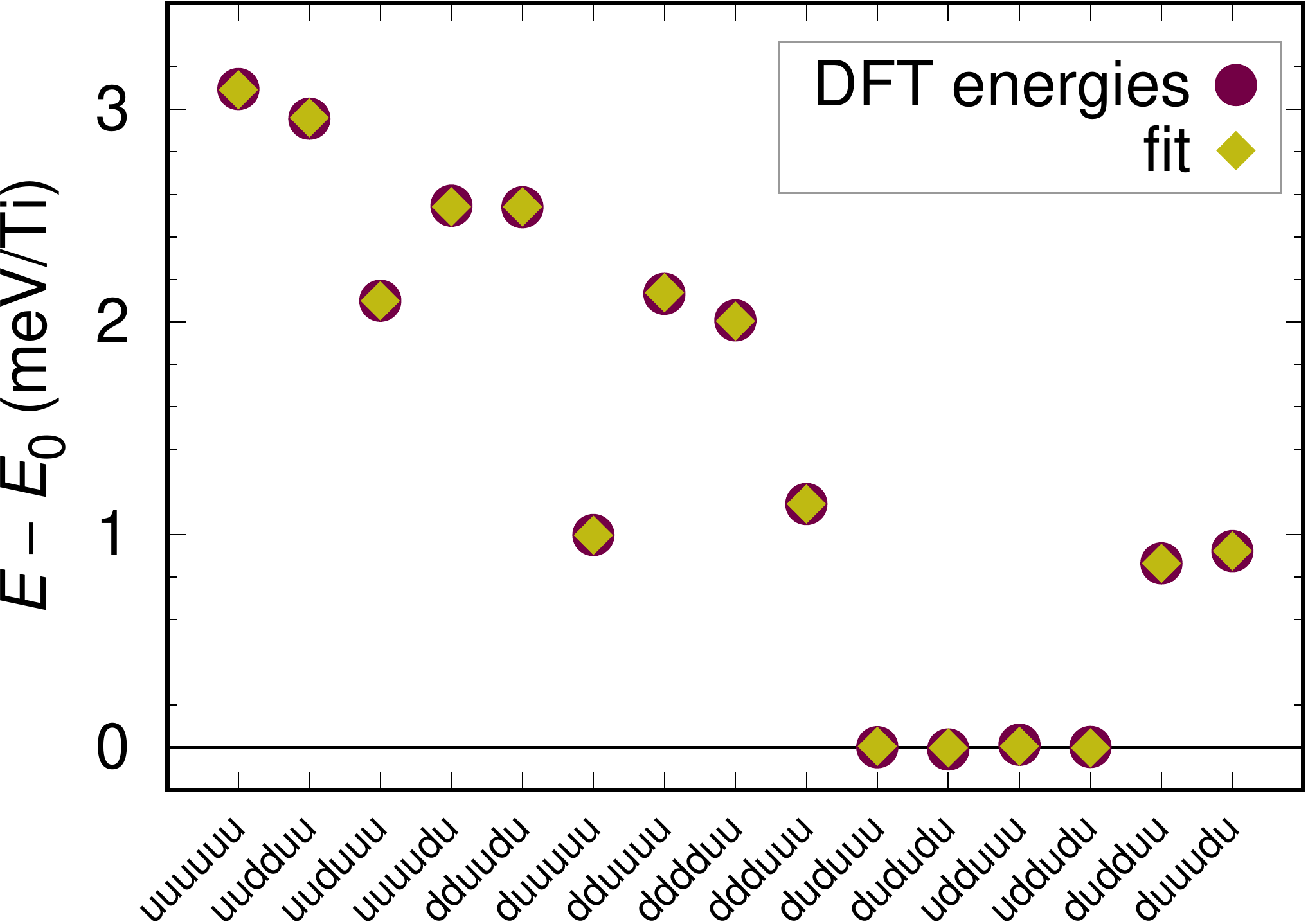}
    \caption{Comparison between GGA+U, $U=2.5$~eV, $J=0.64$~eV total energies (per Ti$^{3+}$) for {\rbna} and the energies calculated from the Heisenberg Hamiltonian, Eq.~\eqref{eq:H} with the six exchange couplings. }\label{fig:config}
\end{figure}

The method to extract Heisenberg Hamiltonian parameters from density functional theory calculations employed here is the energy mapping technique. We reduce the symmetry of the structures of all three compounds {\rbna}, {\csna} and {\csk} from $P\,2_1/m$ (No. 11) to $P\,1$ (No. 1) which makes all six Ti$^{3+}$ ions in the unit cell inequivalent. This allows for 64 different spin configurations, 15 of which have unique energies. Figure~\ref{fig:config} shows for one example the typical quality of the fit for the three materials.

\section{Numerical diagonalization}\label{app:B}
\begin{figure}[tb]
\includegraphics[width=\linewidth]{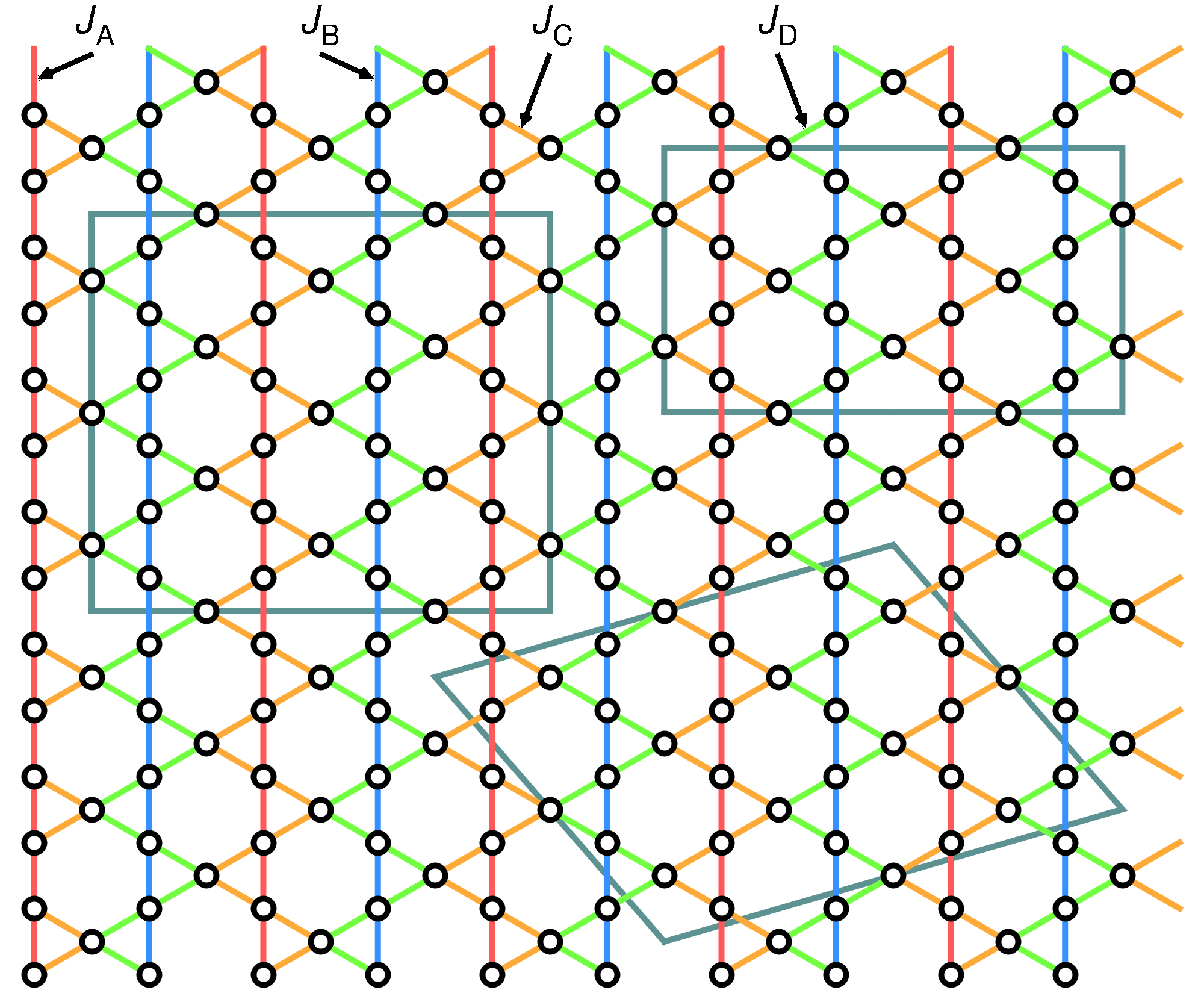}
\caption{(Color online) Illustration of the 24 site, 30 site and 36 site clusters on the kagome lattice used in the numerical diagonalization calculations.}
\label{fig:cluster}
\end{figure}

 Numerical-diagonalization is an unbiased, numerically exact method which gives reliable information for a given Hamiltonian on finite-size clusters of $N=24$, 30, and 36. The shapes of the chosen finite-size clusters are illustrated  in Figure~\ref{fig:cluster}. We employ the periodic boundary conditions in each cluster. 

Diagonalization calculations directly give the lowest energy of the Heisenberg Hamiltonian, which is denoted by $E(N,M_z)$ for each subspace of a given $M_z$. To obtain the step-like magnetization process when the Zeeman term $-2\mu_{\rm B} H \sum_{i} S_i^z$ is added to the Hamiltonian \eqref{eq:H}, we use the relationship $ 2\mu_{\rm B } H=E(N,M_{z}+1) - E(N,M_{z}) $ 
which determines the magnetic field $H$ for the occurrence of the magnetization increase from $M_{z}$ to $M_{z}+1$. Note here that we assume the isotropy of the system and that the normalized magnetization per Ti$^{3+}$ ion, $M$, given by $M_{z}/(NS)$ is used in Fig.~\ref{fig:magstep}.

\end{document}